\documentclass[12pt]{iopart}
\usepackage{amsfonts}
\begin{document}

\title[Induced charge on a graphitic nanocone
at finite temperature]{On the possible induced charge on a graphitic
nanocone at finite temperature}

\author{Yu A Sitenko$^{1,2}$ and N D Vlasii$^{1,2}$}
\address{$^{1}$ Bogolyubov Institute for Theoretical Physics,
National Academy of Sciences, \\ 14-b Metrologichna Str., Kyiv,
03680, Ukraine}
\address{$^{2}$ Physics Department, National Taras Shevchenko University of
Kyiv, \\ 2 Academician Glushkov Ave., Kyiv, 03127, Ukraine}
\ead{yusitenko@bitp.kiev.ua}

\begin{abstract}
Electronic excitations in a graphitic monolayer (graphene) in the
long-wavelength approximation are characterized by the linear
dispersion law, representing a unique example of the really
two-dimensional "ultrarelativistic" fermionic system which in the
presence of topological defects possesses rather unusual properties.
A disclination that rolls up a graphitic sheet into a nanocone is
described by a pointlike pseudomagnetic vortex at the apex of the
cone, and the flux of the vortex is related to the deficit angle
of the conical surface. A general theory of planar relativistic
fermionic systems in the singular vortex background is employed, and
we derive the analytical expression for the charge which is induced
at finite temperature on some graphitic nanocones.
\end{abstract}

\submitto{\JPA}
\maketitle

\section{Introduction}

Carbon nanostructures are intensively studied both experimentally
and theoretically, in view of their expected important implications
for the development of electronic devices, flat panel displays,
nanoswitches, etc. (see, e.g. \cite{Sai}). The recent synthesis of
stricty two-dimensional atomic crystals of carbon (monolayers of
graphite -- graphene) \cite{Nov4,Nov5} is promising a wealth of new
phenomena and applications in technology. The observation of
anomalous transport properties, and, most exciting, the discovery
of substantial field effect and magnetism at room temperature
allows one to envisage graphene as a reasonable replacement
of nanotubes in electronic applications \cite{Ge,Kat}.

In the case of isolated graphene, the electronic states near the
Fermi level can be described in a simple manner \cite{Wa}. By
symmetry, the lower and upper bands touch at the corners of the
hexagonal Brillouin zone. In the vicinity of these points, the
dispersion relation is isotropic and linear, and the density of
states at the Fermi level is strictly zero. Using the tight-binding
approximation for the nearest neighbour interaction in the honeycomb
lattice, an effective long-wavelength description of electronic
states in graphene can be written in terms of a continuum model
which is based on the Dirac--Weyl equation for massless electrons
in 2+1-dimensional space-time with the role of speed of light $c$
played by Fermi velocity $v\approx c/300$ \cite{Di,Sem}. The
one-particle Hamiltonian operator of the model takes the form
\begin{equation}\label{1}
    H=-i\hbar v(\alpha^1\partial_x+\alpha^2\partial_y),
\end{equation}
where the $4\times4$ matrices can be chosen in the form \cite{Si7}
\begin{equation}\label{2}
    \alpha^1=-\left(
               \begin{array}{cc}
                 \sigma^2 & 0 \\
                 0 & \sigma^2 \\
               \end{array}
             \right),\qquad \alpha^2=\left(
                                       \begin{array}{cc}
                                         \sigma^1 & 0 \\
                                         0 & -\sigma^1 \\
                                       \end{array}
                                     \right),
\end{equation}
$\sigma^j (j=1,2,3)$ are the Pauli matrices. The one-particle electronic
wave function possesses 4 components:
\begin{equation}\label{3}
\psi=\left(\psi_{A+},\,\psi_{B+},\,\psi_{A-},\,\psi_{B-}\right)^T,
\end{equation}
where subscripts $A$ and $B$ correspond to two sublattices of the
honeycomb lattice and subscripts $+$ and $-$ correspond to two
inequivalent Fermi points. The linear dispersion law, $E=\pm \hbar
v|\bf{k}|$, results in the density of states
\begin{equation}\label{4}
    \tau(E)=\frac{S|E|}{\pi\hbar^2v^2},
\end{equation}
which is the density of states of two-dimensional ultrarelativistic
Fermi gas (here $S$ is the area of the graphene sample).
Consequently, the thermal average charge of electronic excitations
in graphene,
\begin{equation}\label{5}
    Q(T)=-\frac e2\int\limits_{-\infty}^{\infty}dE\,\tau(E)\,{\rm tanh}\left(\frac12\beta E\right),\,\,\,\,
    \beta=(k_BT)^{-1},
\end{equation}
is evidently zero, because Eq.(4) is even in energy ($k_B$ is the
Boltzmann constant).

Topological defects in graphene are disclinations in the honeycomb
lattice, resulting from the substitution of a hexagon by, say, a
pentagon or a heptagon; such a disclination rolls up the graphitic
sheet into a cone. More generally, a hexagon is substituted by a
polygon with $6-N_d$ sides, where $N_d$ is an integer which is
smaller than 6. Polygons with $N_d>0$ ($N_d<0$) induce locally
positive (negative) curvature, whereas the graphitic sheet is flat
away from the defect, as is the conical surface away from the apex.
In the case of nanocones with $N_d>0$, the value of $N_d$ counts the
number of sectors of the value of $\pi/3$ which are removed from the
graphitic sheet. If $N_d<0$, then $-N_d$ counts the number of such
sectors which are inserted into the graphene sheet. Certainly,
polygonal defects with $N_d>1$ and $N_d<-1$ are mathematical
abstractions, as are cones with a pointlike apex. In reality, the
defects are smoothed, and $N_d>0$ counts the number of the pentagonal
defects which are tightly clustered producing a conical shape; such
nanocones were observed experimentally \cite{Kri}. Theory predicts
also an infinite series of the saddle-like nanocones with $-N_d$
counting the number of the heptagonal defects which are clustered
in their central regions. Saddle-like cones serve as an element
which is necessary for joining parts of carbon nanotubes of
differing radii and for creating
schwarzite \cite{Park}, a structure appearing in many forms of
carbon nanofoam \cite{Rode}.  As it was shown by using
molecular-dynamics simulations \cite{Iha}, in the case of
$N_d\leq-4$, a surface with a polygonal defect is more stable than a
similarly shaped surface containing a multiple number of heptagons;
a screw dislocation can be presented as the $N_d\rightarrow -\infty$
limit of a $6-N_d$-gonal defect.

In the present paper we shall consider the influence of topological
defects in graphene on its electronic properties at finite
temperature\footnote{The case of zero temperature was discussed
earlier in works \cite{Lam,Char,Osi,Voz1,Voz2,Si7}.}. The effects of
the variation of the bond length or the mixing of $\pi$- with
$\sigma$-orbitals caused by curvature of the lattice surface are
neglected, and our consideration, focusing on global aspects of
coordination of carbon atoms, is based on the long-wavelength
continuum model originating in the tight-binding approximation for
the nearest neighbour interactions. We employ a general theory of
planar relativistic fermionic systems in the singular vortex
background \cite{Sit6, Sit7}; its version for the case of massless
fermions was elaborated in \cite{Si9,Sit9,Si0}.

\section{Hamiltonian, resolvent and density of states on a graphitic nanocone}

The Dirac--Weyl Hamiltonian for electronic excitations on a
graphitic nanocone with a pointlike apex takes the form
\begin{equation}\label{6}
    H=-i\hbar v\left\{\alpha^1\partial_r+\alpha^2r^{-1}\left[(1-\eta)^{-1}\partial_\varphi-i\Sigma\right]\right\},
\end{equation}
where
\begin{equation}\label{7}
    \Sigma=\frac1{2i}\alpha^1\alpha^2=\frac12\left(
  \begin{array}{cc}
   \sigma^3 & 0 \\
    0 &-\sigma^3  \\
  \end{array}
\right)
\end{equation}
is the pseudospin, $r$ and $\varphi$ are the polar coordinates
centred at the apex of the cone with metric
\begin{equation}\label{23}
    g_{rr}=1,\quad g_{\varphi\varphi}=(1-\eta)^2r^2,
\end{equation}
and
\begin{equation}\label{8}
    \eta=N_d/6.
\end{equation}
The electronic wave function on a graphitic nanocone obeys the
M\"{o}bius-strip-type condition \cite{Si7}:
\begin{equation}\label{9}
    \psi(r,\,\varphi+2\pi)=-\exp\left(-i\frac\pi2N_dR\right)\psi(r,\,\varphi),
\end{equation}
where
\begin{equation}\label{10}
    R=i\left(
  \begin{array}{cc}
    0 & \sigma^2 \\
    -\sigma^2 & 0 \\
  \end{array}
\right)
\end{equation}
is the operator exchanging sublattices, as well as inequivalent
Fermi points, and commuting with Hamiltonian (6).

By performing singular gauge transformation
\begin{equation}\label{11}
    \psi'=e^{i\Omega}\psi,\quad \Omega=\varphi\frac{N_d}{4}R,
\end{equation}
one gets the wave function obeying usual condition
\begin{equation}\label{12}
    \psi'(r,\,\varphi+2\pi)=-\psi'(r,\,\varphi),
\end{equation}
in the meantime, Hamiltonian (6) is transformed to
\begin{equation}\label{13}
    \!\!\!\!H'\!=\!e^{i\Omega}He^{-i\Omega}\!=\!-i\hbar v\left\{\alpha^1\partial_r+\alpha^2r^{-1}\left[
    (1-\eta)^{-1}\!\left(\partial_\varphi-i\frac{3}{2}\eta R \right)\!-i\Sigma\right]\right\},
\end{equation}
where relation (9) is recalled. Thus, a topological defect in
graphene is presented by a pseudomagnetic vortex with flux
$N_d\,\pi/2$ through the apex of a cone with deficit angle
$N_d\,\pi/3$.

By performing a unitary transformation which diagonalizes $R$, one
gets Hamiltonian in the block-diagonal form
\begin{equation}\label{14}
    H''=\left(
                               \begin{array}{cc}
                                 H_1 & 0 \\
                                 0 & H_{-1} \\
                               \end{array}
                             \right),
\end{equation}
where
\begin{equation}\label{15}
    \!\!\!\!\!H_s=\hbar v\left\{i\sigma^2\partial_r-\sigma^1r^{-1}\left[(1-\eta)^{-1}\left(
    is\partial_\varphi+\frac32\eta\right)+\frac12\sigma^3\right]\right\}, \,\,\, s=\pm1.
\end{equation}
Let us consider the kernel of resolvent (Green's function) of
Hamiltonian (16)
\begin{equation}\label{16}
   \!\!\!\!\!\!\!\!\!\!\! \langle r,\,\varphi|(H_s-\omega)^{-1}|r',\,\varphi'\rangle=\frac 1{2\pi}\sum\limits_{n\in \mathbb{Z}}e^{i
    \left(n+\frac s2\right)(\varphi-\varphi')}\left(
                                                \begin{array}{cc}
                                                  a^{(n)}_{11}(r,\,r') & a^{(n)}_{21}(r,\,r')\\
                                                  a^{(n)}_{12}(r,\,r') & a^{(n)}_{22}(r,\,r')\\
                                                \end{array}
                                              \right),
\end{equation}
where $\mathbb{Z}$ is the set of integer numbers, and $\omega$ is a
complex parameter with dimension of energy. Off-diagonal radial
components are expressed through the diagonal ones, while the latter
satisfy the second-order differential equations. All radial
components behave asymptotically at large distances as outgoing
waves, i.e. as $e^{ikr}(2\pi\sqrt{r})^{-1}$ at $r\rightarrow \infty$
and as $e^{ikr'}(2\pi\sqrt{r'})^{-1}$ at $r'\rightarrow \infty$,
where $k=\sqrt{\omega^2}\,(\hbar v)^{-1}$ and a physical sheet for the
square root is chosen as $0<{\rm Arg}k<\pi$ (${\rm Im}k>0$). In the
cases of $N_d=3,\,4,\,5$ all radial components are regular at small
distances, i.e. at $r\rightarrow 0$ and at $r'\rightarrow 0$. In all
other cases there are radial components that are irregular at $r=0$
and $r'=0$ for some values of $n$. This is related to the fact that
in these cases Hamiltonian (16) is not essentially self-adjoint and
is characterized by the nonzero deficiency index. According to the
Weyl--von Neumann theory of self-adjoint operators (see, e.g. \cite{Alb}),
a procedure of the self-adjoint extension is implemented
yielding the condition for the irregular radial components, which
depends on the set of self-adjoint extension parameters. It should
be emphasized that only irregular radial components can produce a
piece in the density of states, which is odd in energy. Thus, if
charge (5) is nonvanishing, then it depends on the set of
self-adjoint extension parameters.

Let us consider the possibility of charge generation in the cases of
$N_d=2,\,1,\,-1,\,-2,\,-3,\,-6$, when Hamiltonian (16) is
characterized by deficiency index $(1,\,1)$, and there is only one
self-adjoint extension parameter -- $\Theta$. The irregular diagonal
radial components are given by expressions
\begin{eqnarray}\label{17}
a_{11}^{(n_c)}(r;r')= \frac{i\pi}{2(1-\eta)}\frac{\omega}{(\sin\nu_\omega+\cos\nu_\omega
e^{iF\pi})}\times \nonumber \\ \times \left\{\theta(r-r')H^{(1)}_{-F}(k r) [\sin\nu_\omega
J_{-F}(k r')+\cos\nu_\omega J_{F}(k r')]+\right. \nonumber \\ \left.+\theta(r'\!-\!r)[\sin\nu_\omega
J_{-F}(k r)+\cos\nu_\omega J_{F}(k r)]H^{(1)}_{-F}(k r') \right\}\!,
\end{eqnarray}
\begin{eqnarray}\label{18}
a_{22}^{(n_c)}(r;r')=\frac{i\pi}{2(1-\eta)}\frac{\omega}{(\sin\nu_\omega+\cos\nu_\omega
e^{iF\pi})}\times \nonumber \\ \times \left\{\theta(r-r')H^{(1)}_{1-F}(k r)
[\sin\nu_\omega J_{1-F}(k r')-\cos\nu_\omega J_{-1+F}(k r')]+\right. \nonumber
\\ \left.+\theta(r'-r)[\sin\nu_\omega J_{1-F}(k r)-\cos\nu_\omega
J_{-1+F}(k r)]H^{(1)}_{1-F}(k r') \right\}\,,
\end{eqnarray}
where
\begin{equation}\label{19}
    n_c=\left\{\begin{array}{cl}
                 \frac{s}{2}[{\rm sgn} (N_d)-1] & N_d=2,\,1,\,-1,\,-2,\,-3, \\
                 -2s, & N_d=-6,
               \end{array}\right.
\end{equation}
\begin{equation}\label{20}
F=\left\{\begin{array}{cl}
             [3-3{\rm sgn}(N_d)+N_d](6-N_d)^{-1}, & N_d=2,\,1,\,-1,\,-2,\,-3, \\
             1/2,&  N_d=-6,
           \end{array}
\right.
\end{equation}
$J_\mu(u)$ is the Bessel function of order $\mu$, $H^{(1)}_\mu(u)$ is
the first-kind Hankel function of order $\mu$,
$$
\theta(u)=\left\{\begin{array}{cc}
  1, & u>0 \\
  0, & u<0
\end{array}\right\},
$$
parameter $\nu_\omega$ is related to $\Theta$ in the following way
\begin{equation}\label{21}
{\rm{tan}}\nu_\omega=\frac{\hbar v k^{2F}}{\omega}\,\left(\frac{Mv}{\hbar}\right)^{1-2F}
\tan\left(\frac\Theta2+\frac\pi4\right)\,,
\end{equation}
$M$ is the scale-invariance-breaking parameter of dimension of mass. The
diagonal radial components for $n\neq n_c$ are regular.

Computing the contribution of the irregular radial components to the
functional trace of the resolvent, we get
$$
    \left[{\rm Tr}(H_s-\omega)^{-1}\right]_{\rm{irreg}}=(1-\eta)\int\limits_{0}^{\infty}dr\,r\,
    \left[a_{11}^{(n_c)}(r,r)+a_{22}^{(n_c)}(r,r)\right]=
$$
\begin{equation}\label{22}
    =\frac{2F-1}{\omega(e^{-iF\pi}\rm{tan}\nu_{\omega}+1)}\, ,
\end{equation}
where ${\rm Tr}\ldots =\int d^2 x \sqrt{g} \, {\rm
tr}\langle{\bf x}|\ldots |{\bf x}\rangle$ and ${\rm tr}$ denotes the
trace over pseudospinor indices only. Using the relation between the
resolvent trace and the density of states,
\begin{equation}\label{23}
    \tau(E)=\frac{1}{2\pi i}\left[{\rm Tr}(H-E-i0)^{-1}-{\rm Tr}(H-E+i0)^{-1}\right],
\end{equation}
we get the expression for the piece of the density of states, which
is odd in energy:
$$
\tau_{{\rm odd}}(E)=\frac{2(2F-1)\sin(F\pi)}{\pi E}\times
$$
\begin{equation}\label{24}
\times\frac{\left(\frac{|E|}{Mv^2}\right)^{2F-1}
    \tan\left(\frac \Theta 2+\frac\pi 4\right)+\left(\frac{|E|}{Mv^2}\right)^{1-2F}
    \cot\left(\frac \Theta 2+\frac\pi 4\right)}{\left(\frac{|E|}{Mv^2}\right)^{2(2F-1)}
    \tan^2\left(\frac \Theta 2+\frac\pi 4\right)\!-\!2\cos(2F\pi)\!+\!\left(\frac{|E|}{Mv^2}\right)^{2(1-2F)}\cot^2\left(\frac
    \Theta 2+\frac \pi 4\right)},
\end{equation}
where the summation over $s=\pm 1$ is performed. Although quantity
(25) is negligible as compared to the ideal gas contribution (4),
lacking the factor of the sample area, it provides the generation of
nonzero charge on some graphitic nanocones.

\section{Induced charge}

Substituting (25) into (5), we get the thermal average of charge,
\begin{eqnarray}\label{25}
  Q(T)=2e(1-2F)\frac{\sin(F\pi)}{\pi}\int\limits_{0}^{\infty}\frac{du}{u}\tanh
  \left(\frac 12u\beta Mv^2\right) \times \nonumber \\
   \times \frac{u^{2F-1}\tan\left(\frac \Theta 2+\frac \pi 4\right)+u^{1-2F}{\rm cot}
   \left(\frac \Theta 2+\frac \pi 4\right)}{u^{2(2F-1)}\tan^2\left(\frac \Theta 2+\frac \pi 4\right)
   -2\cos(2F\pi)+u^{2(1-2F)}{\rm cot}^2\left(\frac \Theta 2+\frac \pi 4\right)}\,.
\end{eqnarray}
Taking into account relation (24), one can present thermal average
(5) in the form
\begin{equation}
Q(T)=-\frac{e}{2}\int\limits_{C}\frac{d\omega}{2\pi i}
{\rm Tr}(H-\omega)^{-1}\tanh(\frac{1}{2}\beta\omega),
\end{equation}
where $C$ is the contour consisting of two collinear straight lines,
$(-\infty+i0,\,+\infty+i0)$ and $(+\infty-i0,\,-\infty-i0)$, in the
complex $\omega$-plane. By deforming contour $C$ to encircle poles
of the hyperbolic tangential function on the imaginary axis, one
gets
\begin{equation}\label{27}
    Q(T)=-\frac e\beta\sum\limits_{m\in \mathbb{Z}}{\rm Tr}(H-i\omega_m)^{-1},
\end{equation}
where $\omega_m=(2m+1)\pi/\beta$. Using (23) and summing over $s=\pm
1$, we get
\begin{eqnarray}\label{28}
Q(T)=\frac{4e}{\beta Mv^2}(1-2F)\times \nonumber \\ \!\!\!\!\!\!\!\!\!\!\!\!\!\!
\!\!\!\!\!\!\!\!\!\!\!\!\!\!\!\!\!\!\!\!\!\times
\sum\limits_{\begin{array}{c}
                                                  m\in \mathbb{Z} \\
                                                  m\geq 0
                                                \end{array}
}\left\{\left[\frac{(2m+1)\pi}{\beta Mv^2}\right]^{2F}{\rm tan}\left(\frac \Theta 2+
\frac \pi 4\right)+
\left[\frac{(2m+1)\pi}{\beta Mv^2}\right]^{2(1-F)}{\rm cot}\left(\frac \Theta 2+
\frac \pi 4\right)\right\}^{-1}
\end{eqnarray}
Representation (26) can be regarded as the Sommerfeld--Watson
transform of the infinite sum representation (29).

Thus, the charge is zero for graphitic nanocones with
$N_d=2,\,-2,\,-6$, since in these cases one has $F=1/2$, see (21).
Note that one has: $F=1/5$ for $N_d=1$, $F=5/7$ for $N_d=-1$, and
$F=1/3$ for $N_d=-3$. Therefore the charge, if any, for the
one-pentagon ($N_d=1$) defect is of opposite sign to that for the
one-heptagon ($N_d=-1$) defect and is of the same sign to that for
the three-heptagon ($N_d=-3$) defect.

In the high-temperature limit $(\beta\rightarrow 0)$ the charge
tends to zero as inverse power of temperature:
\begin{eqnarray}\label{d16}
Q(T\rightarrow\infty)=\frac{4e}{\pi}(1-2F)\times \nonumber \\ \!\!\!\!\!\!\!\!\!\!\!\!\!\!
\!\!\!\!\!\!\!\!\!\!\!\!\!\!\!\!\!\!\!\!\!\times
\left\{\begin{array}{lr}{\displaystyle (1-2^{-2+2F})\zeta(2-2F)
\tan\left(\frac\Theta2+\frac\pi4\right)\left(\frac{\beta Mv^2}{\pi}\right)^{1-2F}},
&{\displaystyle0<F<\frac12}\\ {\displaystyle (1-2^{-2F})\zeta(2F)
\cot\left(\frac\Theta2+\frac\pi4\right)\left(\frac{\beta Mv^2}{\pi}\right)^{2F-1}}
,&{\displaystyle\frac12<F<1}
\end{array}\right.,
\end{eqnarray}
where $\zeta(u)$ is the Riemann zeta function (see, e.g.
\cite{Jorg}). In the zero-temperature limit the charge tends to
finite value \cite{Si7}:
\begin{equation}\label{29}
    Q(0)= e\,{\rm sgn}_0[(1-2F)\cos\Theta],
\end{equation}
where
$$
{\rm sgn}_0(u)=\left\{\begin{array}{cc}
                        \theta(u)-\theta(-u), & u\neq 0 \\
                        0, & u=0
                      \end{array}
\right\}.
$$
Thus, there is a remarkable equation relating the ground state charges
which are induced by the one-pentagon, one-heptagon and
three-heptagon defects:
\begin{equation}\label{30}
Q(0)|_{N_d=1}=-Q(0)|_{N_d=-1}=Q(0)|_{N_d=-3}.
\end{equation}
As it is shown in \cite{Si7}, the induced ground state charge is
accumulated around a defect. This suggests that the same happens for
the temperature-induced charge as well, and the local density of
states is modified in essential way in the vicinity of a defect.

We conclude that the theory predicts the dependence of the induced
charge on the self-adjoint extension parameter in the cases of
graphitic nanocones with one pentagon, one heptagon and three
heptagons at the apex. This is due to the fact that the density of
electronic states attains an additional piece which is odd in
energy, see (25). Actually, there are three possibilities:
$\cos\Theta>0$, $\cos\Theta<0$, and $\cos\Theta=0$. The question of
which of the possibilities is realized has to be answered by future
experimental measurements.

\ack{}

The work of Yu.A.S. was supported by grant No. 10/07-N "Nanostructure systems,
nanomaterials, nanotechnologies" of the National Academy of Sciences
of Ukraine and grant No. 05-1000008-7865 of the INTAS. The work of N.D.V.
was supported by the Young Scientist Fellowship grant (No. 05-109-5333) of
the INTAS. We acknowledge the partial support of the Swiss National
Science Foundation under the SCOPES project No. IB7320-110848.

\end{document}